# Arctic Sea Ice and the Mean Temperature of the Northern Hemisphere

*Alfred Laubereau and Hristo Iglev*

Physik-Department E11, Technische Universität München, James-Franck-Strasse, D-85748 Garching, Germany

ABSTRACT. The importance of snow cover and ice extent in the Northern Hemisphere was recognized by various authors leading to a positive feedback of surface reflectivity on climate. In fact, the retreat of Arctic sea ice is accompanied by enhanced solar input in the Arctic region, i.e. a decrease of the terrestrial albedo. We have studied this effect for the past six decades and estimate the corresponding global warming in the northern hemisphere. A simple 1-dimensional model is used that includes the simultaneous increase of the greenhouse gases. Our results indicate that the latter directly cause a temperature rise of only 0.2 K in 1955 to 2015, while a notably larger effect 0.7 ± 0.2 K is found for the loss of Arctic sea ice in the same time. These numbers comprise most of the reported mean temperature rise of 1.2 ± 0.2 K of the northern hemisphere. The origin of the sea-ice retreat is discussed, e.g. internal variability or feedback by the $CO_2$ concentration increase. Our data also suggest a delayed response of the global surface temperature rise to the loss of sea ice with a time constant of approximately 10 to 20 years.

**KEYWORDS** Global warming, Arctic sea ice, albedo change

INTRODUCTION:

In the Arctic a tremendous loss of sea ice occurred in the past decades accompanied by a warming larger than the global average known as Arctic amplification (1-6). The trend is continuing and may lead to an ice-free Arctic within the next few decades. The prospect has received great scientific attention and public interest (7-9). Various publications arrived at the conclusion that the ice loss originates only in part from man-made factors. A contribution of 30 to 50 percent was estimated for internal variability (3). The surface temperature of the earth, on the other hand, depends on various natural and anthropogenic factors that affect the solar input and the infrared reemission of the earth. Examples are volcanic activities, oceanic temperature profiles and the concentration of the greenhouse gases (GHG) (10-14). An interesting question in this context is the causal relationship between ice-loss and global warming. In the following we elucidate the role of the polar ice cover for the mean surface temperature $T_{surf}$ of the Northern Hemisphere (NH). We present arguments that the global temperature rise is to a larger extent the direct consequence of the sea-ice retreat than vice versa.

Our discussion is based on reported data of Arctic sea ice as empirical facts. Since 1979 results on the maximum and minimum ice areas are available from satellite observations including the phase shift relative to the seasonal solar input. The data are depicted in Fig. 1 supplemented by linear fits (red points and straight solid lines) (15-17). The ice maxima in winter decline from 14.90 to $13.47 \cdot 10^{12}$ m$^2$ while the September minima drop from 5.59 to $3.07 \cdot 10^{12}$ m$^2$ in the years 1979 to 2015. The loss of the sea ice area is noteworthy. The data of the ice areas are extended back to 1955 (note fitted red lines) by the help of auxiliary information on the sea ice extent (18,19) that are available also for earlier years. The reported annual average sea ice extent is depicted in the figure (blue experimental points and fitted blue lines, right hand ordinate scale). Since the ice extent displays an approximately linear dependence since 1955 the same behavior is taken for the ice area; i.e. the linear slope for the measurements 1979 – 2015 is extrapolated until 1955 in Fig 1. For earlier years, the slope of the ice extent data changes (see dashed blue line) indicating a limit for the extrapolation. As a result, the maximum and minimum ice areas are estimated to be 15.85 and $7.26 \cdot 10^{12}$ m$^2$, respectively, in 1955. The numbers represent a decrease of minimum ice area of approximately 58% in 1955 to 2015 (15% for the winter maxima). Naive extrapolation of the linear slopes to the future would suggest a drop of the summer ice minima below $1 \cdot 10^{12}$ m$^2$ around 2050, i.e. practically ice-free Arctic (20). Taking into account the smaller backscattering of sea water compared to snow-covered ice for incident sun light (factor of 0.5) and averaging over the daily and seasonal variation of the solar input, the net input of solar radiation is estimated to grow by $3.22 \cdot 10^{14}$ W in 1955 to 2015. Relative to the total solar input of the NH ($8.7 \cdot 10^{16}$ W) we arrive at a decrease of the global albedo (0.300) of 1.23% (see appendix).

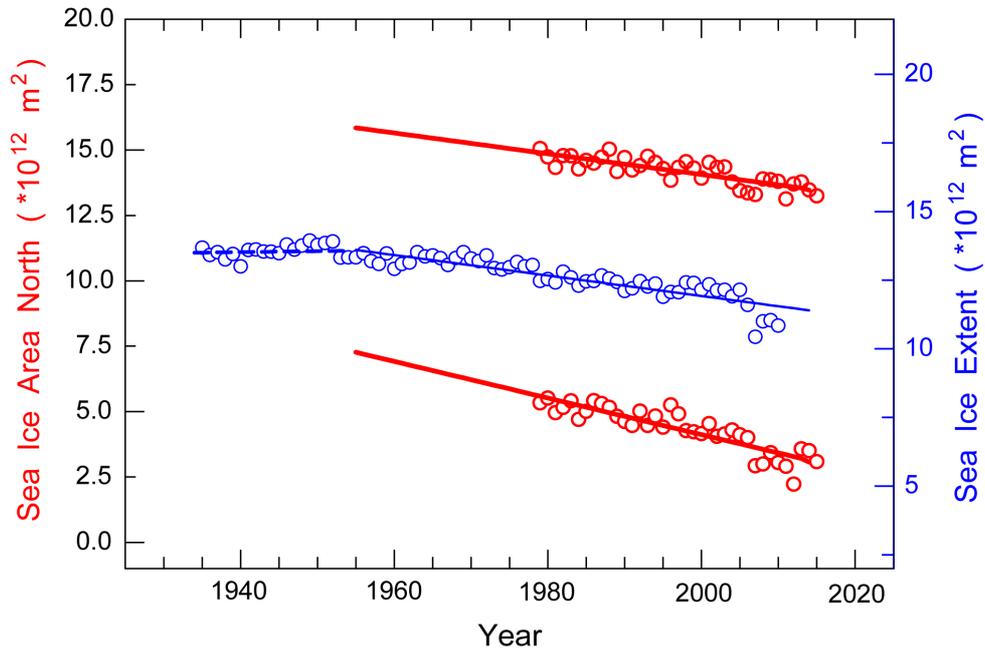

**Figure 1.** Decrease of Arctic sea ice in the years 1955 to 2015 (open red circles and fitted solid red line). Auxiliary data for the average sea ice extent is included (blue data, r.h.s. ordinate scale) with a minor increase earlier to 1955 (dashed blue line) and decay later on. Experimental points taken from literature and calculated lines; see text; after (15-19).

We have developed a simple 1-dimensional model for the causal effect of albedo changes for the mean NH surface temperature including also the greenhouse effect. The model is based on the spectroscopic properties of the gases and several empirical facts. The atmosphere is represented by four layers with mean temperatures $T_j$ ($j = 1 - 4$). The approach is an extension of the well-known 2-layer model for the greenhouse effect (21-23). The surface is represented by a black radiator with intensity $\sigma \cdot T_{surf}^4$ (Stefan-Boltzmann constant $\sigma$) while the atmospheric parts are treated as selective thermal emitters according to the spectral properties. Transport phenomena between the layers are included but are found to have little quantitative effect on the surface warming (23). The temperatures $T_j$ are self-consistently determined from thermal quasi-equilibrium between the solar input and the thermal emissions of the surface and atmospheric layers. Several empirical facts, e.g. the total solar input 343.3 W/m², albedo = 0.300 and the surface temperature $T_{surf}$ = 288.0 K in the year 1880 are used in the computation. The reported concentrations of the greenhouse gases (GHG: $CO_2$, $CH_4$, $N_2O$) including $H_2O$ also enter the calculations (24,25). The model is an improved version of our earlier work (23) taking into account the spectral properties of water vapor; details will be published elsewhere. We consider the 1-dimensional treatment quite satisfactory in the sense of a Taylor expansion of the (unknown) function $\Delta T_{surf}$ depending on a variety of local and geographical parameters and the concentrations of greenhouse gases (GHG), where second and higher order terms are neglected. Retaining only the first order terms of the expansion, the problem is linearized and averaging over the manifold of parameters can be interchanged leading to a 1-D treatment for the albedo and GHG effect.

RESULTS AND DISCUSSION:

Unfortunately, sufficiently precise measurements of the albedo factor do not exist that would allow a comparison of the ice loss with albedo changes. It is interesting to compare the predicted temperature rise of the model with the time evolution of the mean surface temperature of the northern hemisphere. The latter information is available in the literature (26), starting in 1880. Only part of that data are shown in Fig. 2 from 1930 onwards (open black circles). The temperature change of the surface $\Delta T_{surf}$ is arbitrarily set to zero in 1880 (not shown in the Fig). There is no clear trend of the measured temperature variations earlier to 1955. We do not attempt to explain these features. Arguments were reported for the influence of the aerosol concentration in the atmosphere (27). For the subsequent years the well-known dramatic rise of $\Delta T_{surf}$ is shown by more than 1 K.

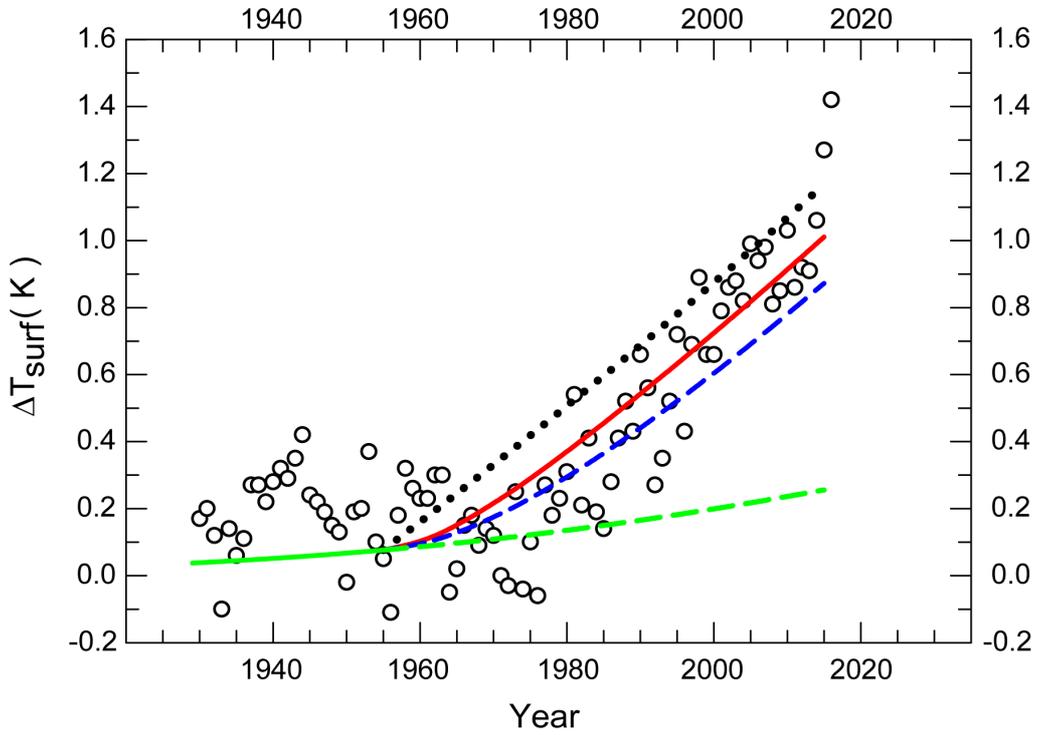

**Figure 2.** Reported change of the mean surface temperature, $\Delta T_{meas}$ (26), and calculated results for the past 60 years. The numbers of reference (26) are upshifted by 0.15 K so that the mean temperature change in 1880 is approximately set to zero. The rise by $\approx$ 1.1 K until 2015 is noteworthy (black circles). For constant albedo = 0.300 the computed temperature change $\Delta T_s$ is depicted, resulting directly from the spectroscopic properties of the GHG including water (green curves). A global warming by 0.26 K is indicated for 2015. For 1955 – 2015 the combined contribution of the GHG and the Arctic ice melting is shown for three situations: (i) dotted black line for instantaneous equilibration within the NH; (ii) delayed equilibration of the mean temperature rise via ice loss with relaxation time $\tau$ = 10 yr (red curve) and (iii) equilibration with $\tau$ = 20 yr (blue line). The fair agreement with the measured temperature rise is noteworthy.

The computed results for the temperature rise by the GHG as derived from our model are shown in Fig. 2 by the green curve. A minor increase of 0.08 K is evaluated until 1955 (see solid green line) with a further rise of 0.18 K for 1955 to 2015 (broken green curve). Comparison with the experimental data (open circles) readily shows that the growing abundance of the GHG is not the direct cause for the warming of the NH surface. The important role of the Arctic sea ice for the temperature rise as revealed by our investigation is indicated by the additional data in the Fig. As mentioned above, the lowering of the albedo factor is estimated to be -0.00368 (-1.23 % of 0.300) if instantaneous response of the NH system to the increase of solar input in the Arctic is assumed. The corresponding mean temperature rise of the NH surface is computed to be 1.17 K until 2015, including the GHG contribution. The temperature increase is shown by the dotted black line in Fig. 2 that differs from the observed data (compare experimental points in the Fig). For a tentative explanation of the deviation we recall that a prompt response of the mean surface temperature of the NH to changes in the Arctic may not be expected.

In fact, the well-known permanent temperature differences between equator and pole area point towards a rather slow equilibration in the NH. As a consequence, we extend our theoretical model and incorporate a delayed response of the global system to changes in the Arctic region with a relaxation time $\tau$ (see appendix). The results for time constants of respectively $\tau = 10$ years (red curve) and 20 years (blue line) are also presented in Fig. 2. The delayed response lowers the increase until 2015 to 1.01 K for $\tau = 10$ yr (0.87 K for 20 yr). The improved agreement with the experimental points for the global temperatures should be noted. It is felt that a time constant of $\approx$ 10 to 20 years is reasonable for the equilibration by heat transfer between the pole area and total NH system. We mention that the warming of the Arctic is reported to be twice the global average during the past three decades. The difference presents evidence for more direct warming of the pole region and delayed equilibration with the NH-system.

The relationship between Arctic warming and sea-ice loss is not well understood (28). Some authors point out that it is partly due to natural reasons (29,30). There is evidence for climate changes in past centuries evolving on time scales of hundreds of years, accompanied by distinct changes of the Arctic sea ice (31-33). Climate-model simulations of the sea-ice retreat differ substantially (34). The loss of sea ice was related to the North Atlantic Oscillation Index (35,36) and the Atlantic Driver (37).

It was also argued that ocean currents appear as the only possible source for the delivering of thermal energy to the Arctic, since the warming started suddenly at the end of a two thousand years cooling period so that an equally sudden cause should be required - not supplied by the greenhouse effect (38). On the other hand, evidence was reported very recently for a 60% contribution of summertime atmospheric circulation to September sea ice loss (3). The authors conclude that 30–50% of the ice retreat may be due to internal variability. Notz and Stroeve reported that approximately 60% of the observed decline from 1979-2011 is externally forced and directly follows anthropogenic $CO_2$ emissions (34,39,). The finding may suggest that the temperature rise attributed in Fig. 2 to the ice retreat partially results from the GHG effect. As to further anthropogenic factors, the present authors wish to point to the precipitation of air pollution that is likely to enhance solar input on sea ice and glaciers (28,40,41).

Several authors discussed the feedback of surface reflectivity on climate (2,42-45). In this context, it is interesting to compare our results with data of Flanner *et al*. (2), who quantified the albedo feedback of snow in the northern hemisphere (NH) between 1979 and 2008. Their result for the NH cryosphere albedo feedback was 0.62 (0.3 -- 1.1) W/(m$^2$ K), larger than estimated from various climate models. Our values for the albedo change of sea ice and global warming are respectively -0.0018 and 0.44 K for this period leading to a larger value of 1.38 W/(m$^2$ K). The difference may be due to the fact that the NH sea ice is declining faster than models simulate (2).

In conclusion, we wish to say that we have studied the causal influence of the shrinking Arctic sea ice and of the growing greenhouse gases on the global warming in the northern hemisphere. Our simple 1-dimensional model appears reliable for semi-quantitative estimates with an accuracy of ±10 to 20%. The estimate of the albedo decrease by the loss of sea ice may be accurate within ±10%. The origin of the sea ice retreat is not fully understood so that the resulting global warming may include to some extent an amplification mechanism of the greenhouse effect. Other factors discussed in the literature are also important (30,46,47). Our investigation reveals a dominant, direct effect of the sea-ice retreat for the global temperature rise that amounts to $\Delta T_{surf} \approx 0.9$ K since 1955.

APPENDIX: Estimate of the Albedo Decrease by the Loss of Artic Sea Ice

The data of Ref. 15 indicate the loss of average sea ice area of $3.29 \cdot 10^{12}$ m$^2$ for the years 1955 to 2015 (compare Fig. 1) and a phase shift of ≈ 81 days for the seasonal ice extent and the sun position between the tropics. Melting replaces snow-covered sea ice by a water surface reducing the backscattering by a factor of ≈ 0.5 (1,48). Averaging over the daily and seasonal changes of the solar input for the respective area around a northern latitude of ≈ 82 degrees (49) and also including an average transmission of 0.49 of the sun input in the atmosphere in the Arctic leads to a reduction factor of 0.0716 for the solar intensity of 1367 W/m$^2$ in the ice area. The decrease of the terrestrial back-reflexion into the universe thus arrives at $\Delta P_{ref} = -3.22 \cdot 10^{14}$ W in 2015 relative to 1955. As compared to the total solar input of the NH of $P_{tot} = 8.73 \cdot 10^{16}$ W, we estimate a relative change of $G = \Delta P_{refl}/P_{tot} = -0.00368$. For instantaneous response of the global system this number would correspond to a relative albedo decrease of $\approx -1.23\%$ in 2015. $G_{rel}$ can be treated as a known function of time according to the ice data of Fig. 1. For delayed response of the total hemisphere leading to an albedo decrease $\Delta albedo(t)$ we introduce the simple relaxation ansatz of Eq. 1 with time constant $\tau$:

$$\frac{d}{dt}\Delta albedo + \frac{\Delta albedo}{\tau} = \frac{G_{rel}(t)}{\tau} \qquad (1)$$

The global albedo $= 0.300 + \Delta albedo(t)$ is time-dependent because of the changing backscattering of the Arctic. Eq. 1 is readily solved for $\Delta albedo(t = 1955) = 0$ and decreases to -0.00305 in 2015 for $\tau = 10$ yr (-0.00249 for $\tau = 20$ yr). Our theoretical model responds quite linearly to small albedo changes with a scaling factor $\Delta T_{surf}/\Delta albedo \approx 247$ K. We mention that the albedo declines by loss of sea ice discussed here is lacking direct experimental verification. Changes of the global albedo of opposite signs ($\pm 0.02$) were reported in the years 1985 – 1997 and 1997–2004 (50,51) (not supported by comparison with the reported temperature data), or that the global albedo remained fairly constant during the past decades (52). We propose that the pronounced rise $\Delta T_{meas}$ of the global surface temperature in Fig. 2 in 1970–2015 gives some experimental support for our estimate of the albedo decrease.

AUTHOR INFORMATION

**Corresponding Authors**

*Email: hristo.iglev@ph.tum.de, alfred.laubereau@ph.tum.de,

**Notes**

The authors declare no competing financial interests.